\documentclass[a4paper]{jpconf}
\usepackage{graphicx}
\begin{document}
\title{Heavy flavours in high-energy nuclear collisions: quenching, flow and correlations}

\author{A. Beraudo$^\dagger$, A. De Pace, M. Monteno, M. Nardi, F. Prino}

\address{INFN, sezione di Torino, Via Pietro Giuria 1, I-10125 Torino}

\ead{$^\dagger$beraudo@to.infn.it}

\begin{abstract}
We present results for the quenching, elliptic flow and azimuthal correlations of heavy flavour particles in high-energy nucleus-nucleus collisions obtained through the POWLANG transport setup, developed in the past to study the propagation of heavy quarks in the Quark-Gluon Plasma and here extended to include a modeling of their hadronization in the presence of a medium. Hadronization is described as occurring via the fragmentation of strings with endpoints given by the heavy (anti-)quark $Q(\overline{Q})$ and a thermal parton $\overline{q}(q)$ from the medium. The flow of the light quarks is shown to affect significantly the $R_{AA}$ and $v_2$ of the final $D$ mesons, leading to a better agreement with the experimental data.
\end{abstract}

Heavy quarks -- indirectly accessible through $D$-mesons, heavy-flavour decay electrons and muons and $J/\psi$'s from $B$ decays -- have been used for a long time as probes of the medium formed in heavy-ion collisions.
In a series of papers~\cite{lange0,lange,lange2} over the last few years we developed a complete setup (referred to as POWLANG) for the study of heavy flavour observables in high-energy nucleus-nucleus collisions, describing the initial hard production of the $Q\overline{Q}$ pairs and the corresponding parton-shower stage through the POWHEG-BOX package~\cite{POW,POWBOX} and addressing the successive evolution in the plasma through the relativistic Langevin equation.
 Here we present a brief summary of our recent efforts aimed at supplementing the above numerical tool by modeling the hadronization of the heavy quarks accounting for the presence of a surrounding medium made of light thermal partons feeling the collective flow of the fluid: a comprehensive exposition can be found in~\cite{lange3}.

In order to simulate the hadronization of heavy quarks in the medium at the end of their propagation in the QGP we proceed as follows. Once a heavy quark $Q$, during its stochastic propagation in the fireball, has reached a fluid cell below the decoupling temperature $T_{\rm dec}$, it is forced to hadronize. One extracts then a light antiquark $\overline{q}_{\rm light}$ (up, down or strange, with relative thermal abundances dictated by the ratio $m/T_{\rm dec}$) from a thermal momentum distribution corresponding to the temperature $T_{\rm dec}$ in the Local Rest Frame (LRF) of the fluid; information on the local fluid four-velocity $u^\mu_{\rm fluid}$ provided by hydrodynamics allows one to boost the momentum of $\overline{q}_{\rm light}$ from the LRF to the laboratory frame. 
A string is then constructed joining the endpoints given by $Q$ and $\overline{q}_{\rm light}$ and is then passed to PYTHIA 6.4~\cite{PYTHIA} to simulate its fragmentation into hadrons (and their final decays). In agreement with PYTHIA, in evaluating their momentum distribution, light quarks are taken as ``dressed'' particles with the effective masses $m_{u/d}\!=\!0.33$ GeV and $m_s\!=\!0.5$ GeV. Concerning $T_{\rm dec}$ the values $0.155$ and $0.17$ GeV are explored. In the following some representative results obtained with the new hadronization procedure are displayed and compared to experimental data (when available).

\begin{figure}[!h]
\begin{center}
\includegraphics[clip,width=0.48\textwidth]{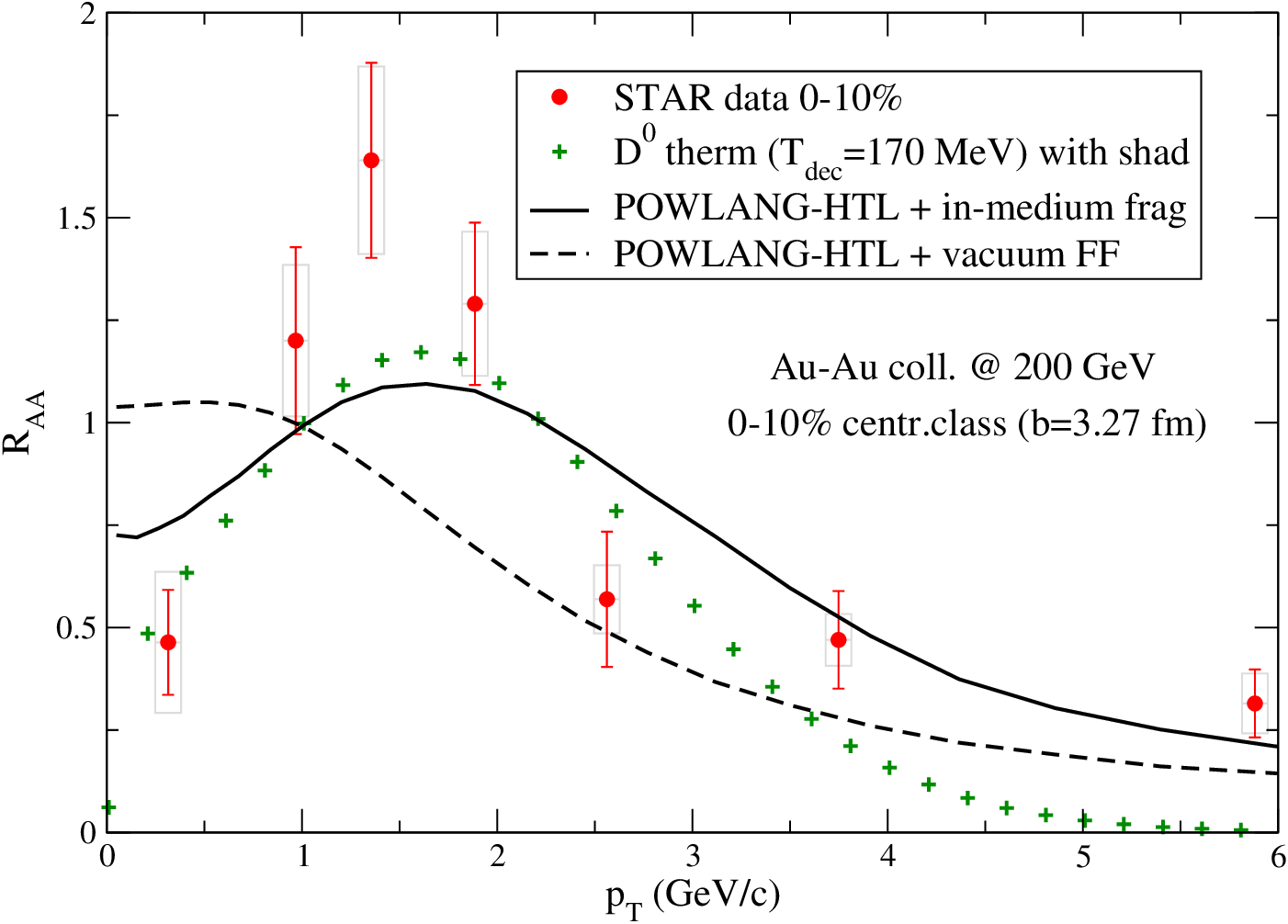}
\includegraphics[clip,width=0.48\textwidth]{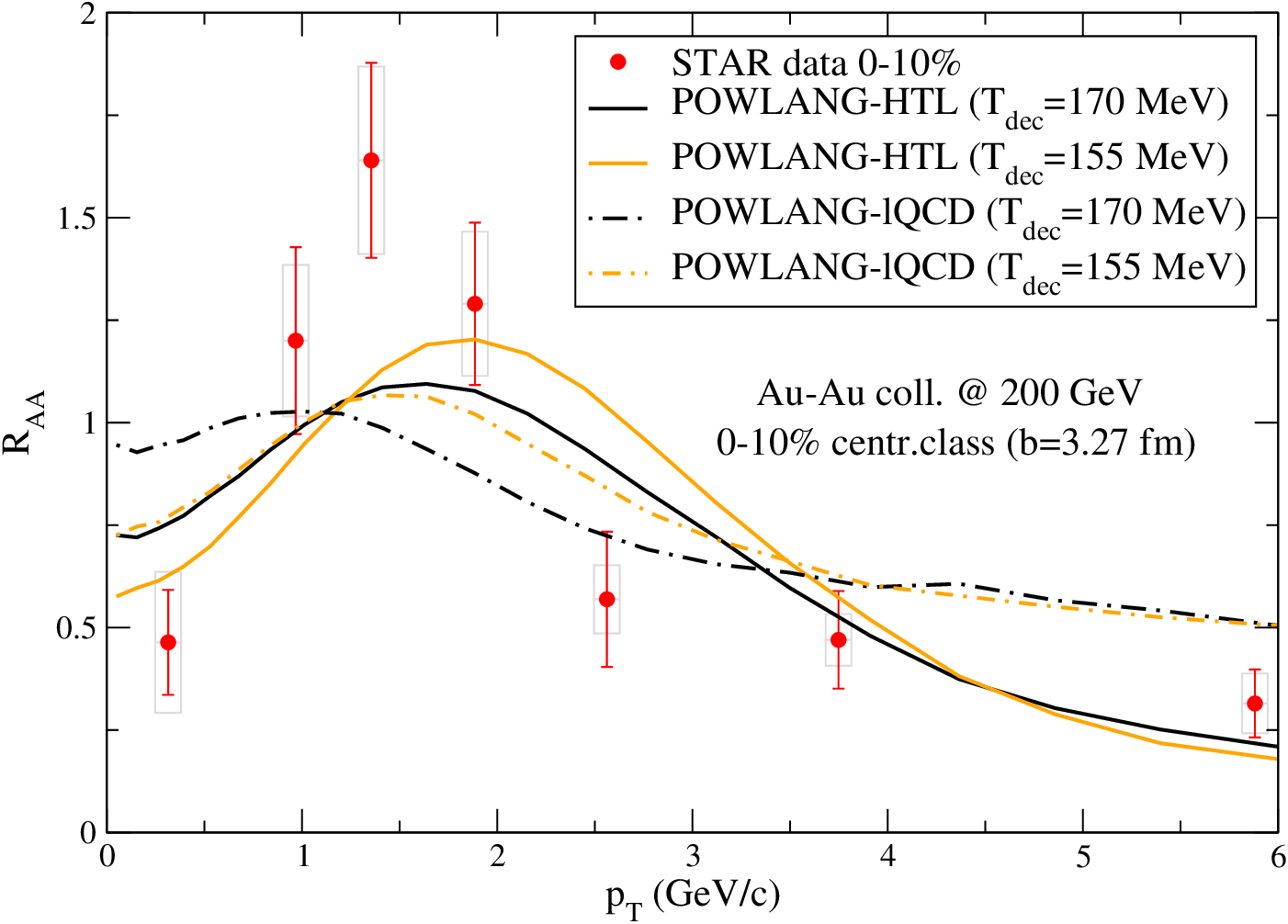}
\caption{$R_{AA}$ of $D^0$ mesons in central Au-Au collisions at $\sqrt{s_{NN}}\!=\!200$ GeV. Theory curves are compared to STAR data~\cite{STARD}. Left panel: POWLANG results obtained with HTL transport coefficients and $T_{\rm dec}\!=\!170$ MeV are displayed, the new in-medium hadronization mechanism leading to a characteristic bump in the $R_{AA}$ due to the radial flow of light quarks; also shown for comparison are the limiting case of full kinetic thermalization of $D$ mesons and the transport result with in-vacuum fragmentation. Right panel: results for two different choices of transport coefficients (weak-coupling HTL, continuous curves, and lattice-QCD, dot-dashed curves) and decoupling temperatures ($T_{\rm dec}\!=\!170$ MeV, in black, and $155$ MeV, in orange) are compared.}\label{fig:RAA_D_RHIC_transp} 
\end{center}
\end{figure}
We start addressing the $D$ meson $R_{AA}$ at RHIC. In Fig.~\ref{fig:RAA_D_RHIC_transp} we show the results of the POWLANG + in-medium fragmentation setup compared to STAR data~\cite{STARD} in central Au-Au collisions at $\sqrt{s_{NN}}\!=\!200$ GeV. One can appreciate how, at variance with the case of vacuum Fragmentation Functions (black dashed curve), due to the additional radial flow inherited from the light thermal parton the theory curves (various choices of transport coefficients and decoupling temperatures are explored) display a bump around $p_T\!\approx\!1.5$ GeV/c in qualitative agreement with the data.

\begin{figure}[!h]
\begin{center}
\includegraphics[clip,width=0.48\textwidth]{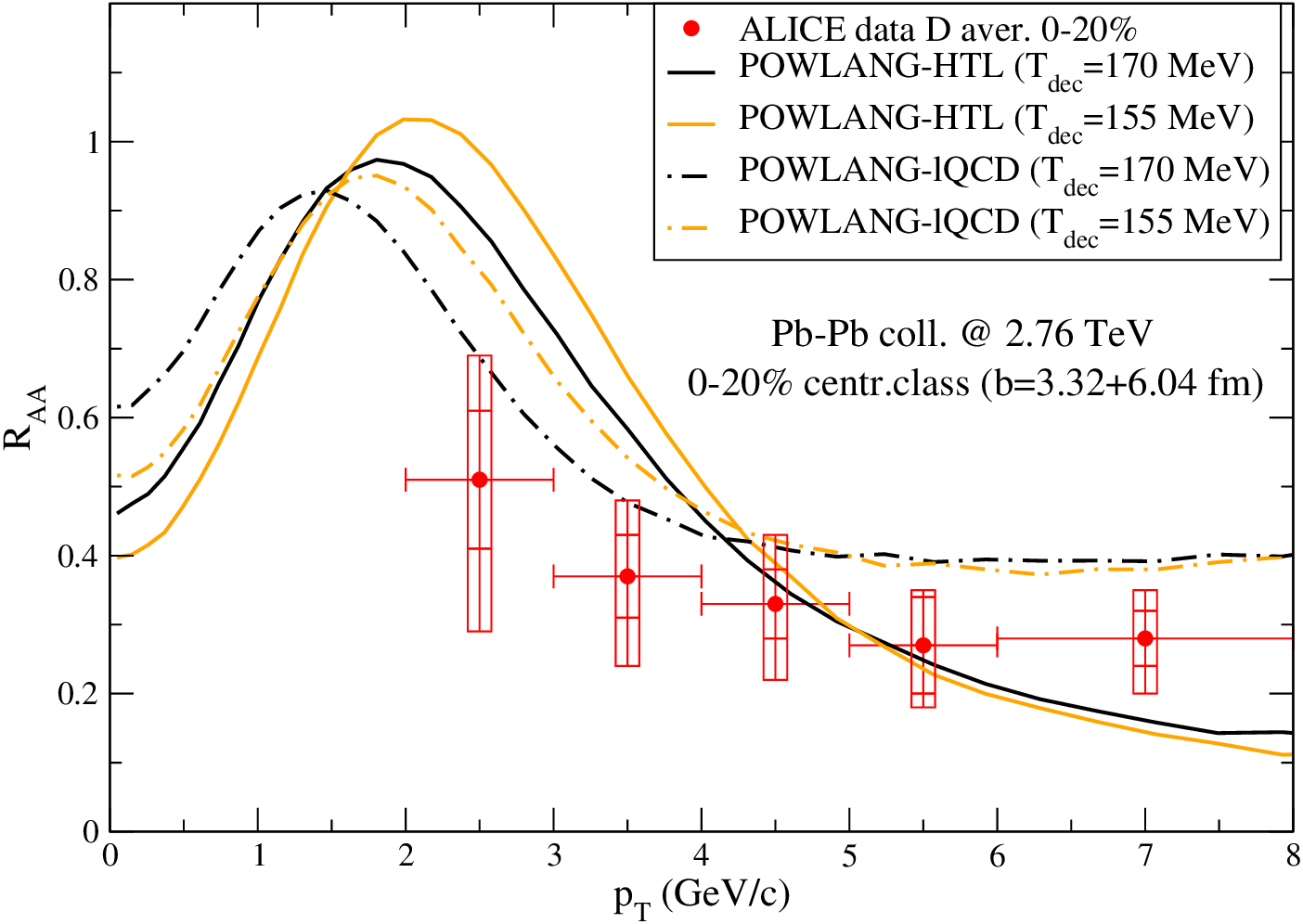}
\includegraphics[clip,width=0.48\textwidth]{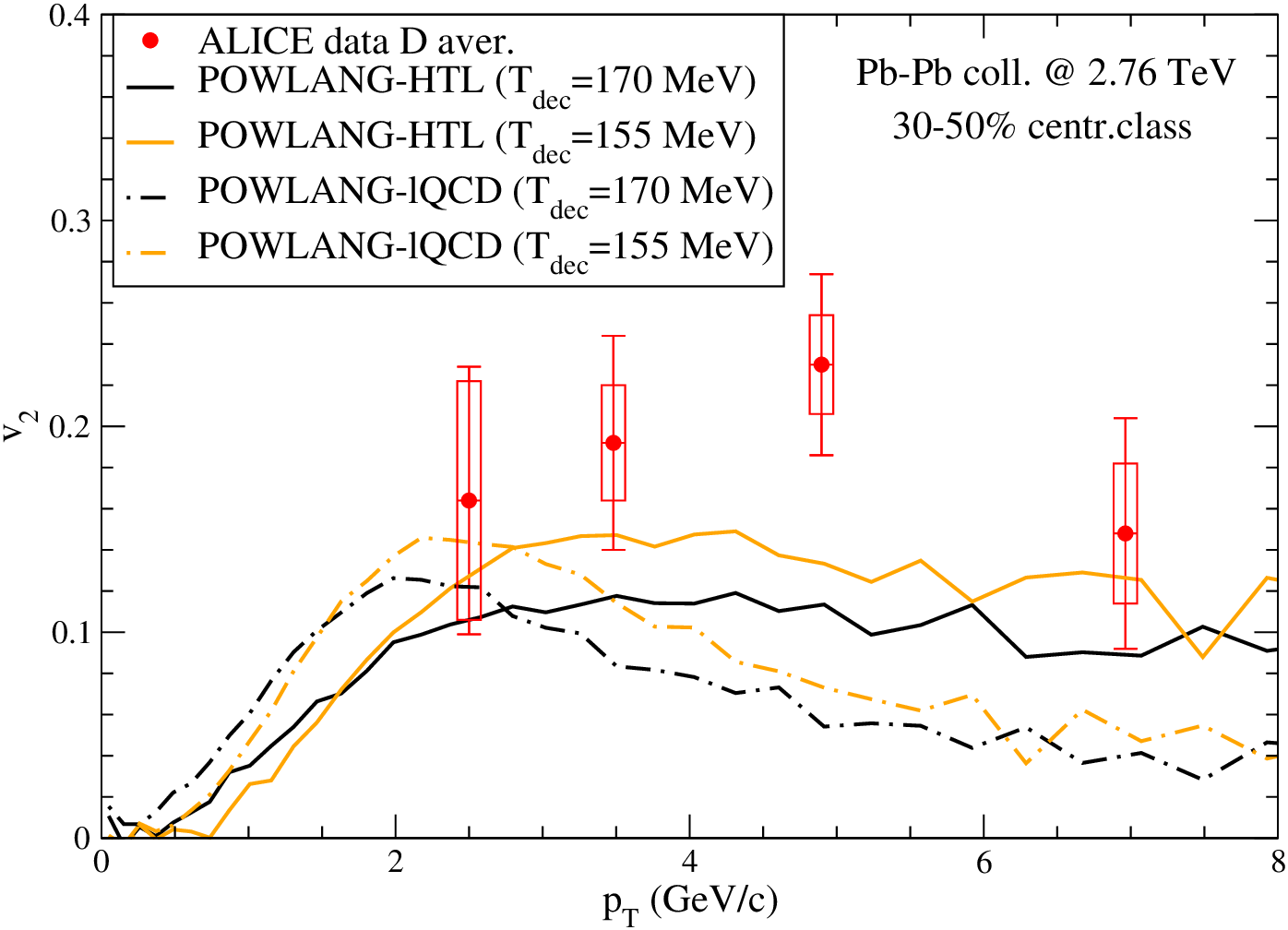}
\caption{The $R_{AA}$ (left panel) and $v_2$ (right panel) of $D$ mesons in central (left) and semi-central (right) Pb-Pb collisions at $\sqrt{s_{NN}}\!=\!2.76$ TeV. POWLANG results with in-medium hadronization and different choices of transport coefficients and decoupling temperatures are compared to ALICE data~\cite{ALICE_DRAA,ALICE_Dv2}.}\label{fig:RAAv2_LHC}
\end{center}
\end{figure}
In Fig.~\ref{fig:RAAv2_LHC} we consider the nuclear modification factor and elliptic flow of $D$ mesons in central (for the $R_{AA}$, left panel) and semi-central (for the $v_2$, right panel) Pb-Pb collisions at $\sqrt{s_{NN}}\!=\!2.76$ at the LHC; theory results are compared to experimental data by the ALICE Collaboration~\cite{ALICE_DRAA,ALICE_Dv2}. Notice that, with the present broader binning and accessible $p_T$ domain, ALICE data for the $D$ meson $R_{AA}$ don't show evidence of the bump from radial flow found by STAR at lower center-of-mass energy. Concerning the  $D$ meson $v_2$, the additional elliptic flow acquired from the light thermal parton picked-up at hadronization moves the theory curves significantly closer to the data with respect to past results with vacuum Fragmentation Functions, which underestimated the experimental findings.  

\begin{figure}[!h]
\begin{center}
\includegraphics[clip,width=0.48\textwidth]{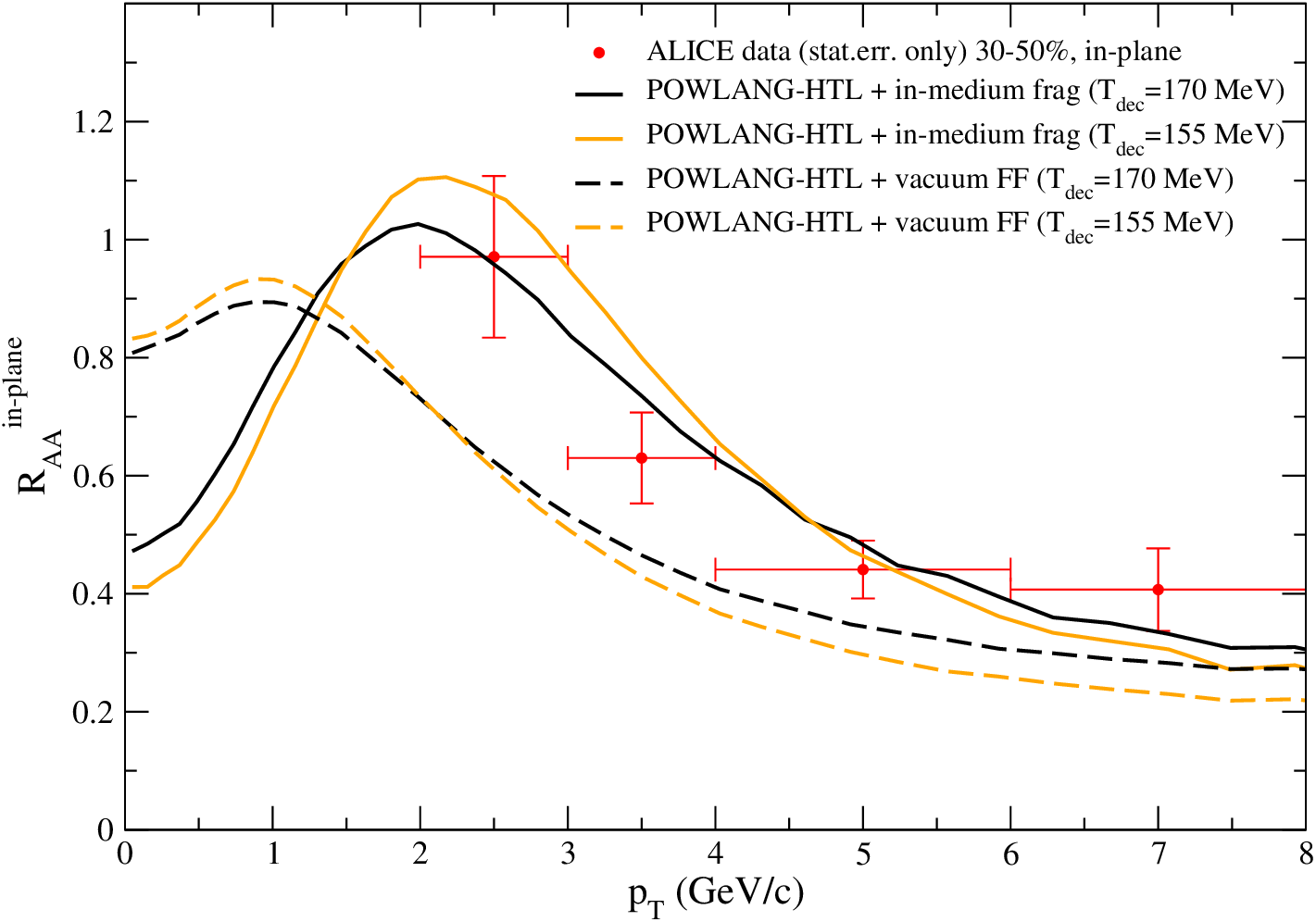}
\includegraphics[clip,width=0.48\textwidth]{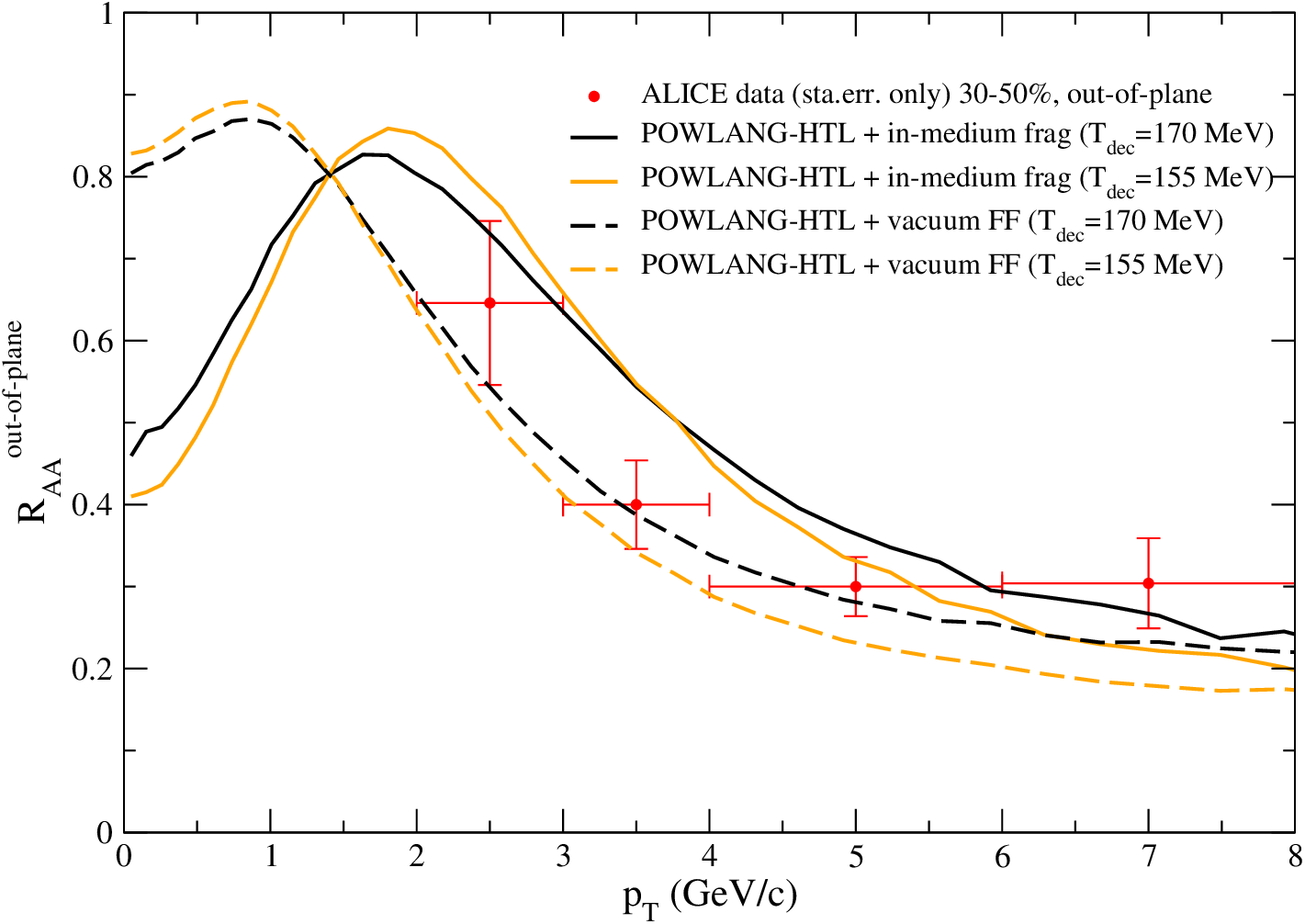}
\caption{$R_{AA}$ in-plane (left panel) and out-of-plane (right panel) of $D$ mesons. ALICE data in the $30\!-\!50\%$ centrality class~\cite{ALICE_reactionplane} are compared to POWLANG results obtained with different hadronization mechanisms (in-medium, solid curves, and vacuum fragmentation, dashed curves) and decoupling temperatures ($T_{\rm dec}\!=\!170$ MeV, black curves, and $155$ MeV, orange curves).}\label{fig:RAA_inout_medvac} 
\end{center}
\end{figure}
A possible way of combining the information contained in the $R_{AA}$ and $v_2$ data (the latter reflecting the collective flow of the medium at low $p_T$ and the path-length dependence of the energy-loss at high $p_T$) consists in studying the azimuthal dependence (with respect to the reaction plane) of the nuclear modification factor of hard particles. The ALICE Collaboration has recently released the data for the in-plane/out-of-plane $R_{AA}$ of $D$ mesons in semi-central (30-50\%) Pb-Pb collisions~\cite{ALICE_reactionplane}. Experimental findings are displayed in Fig.~\ref{fig:RAA_inout_medvac} and compared to the transport predictions of the POWLANG setup: also in this case model results with in-medium fragmentation look in slightly better agreement with the experimental data.  

\begin{figure}[!h]
\begin{center}
\includegraphics[clip,width=0.48\textwidth]{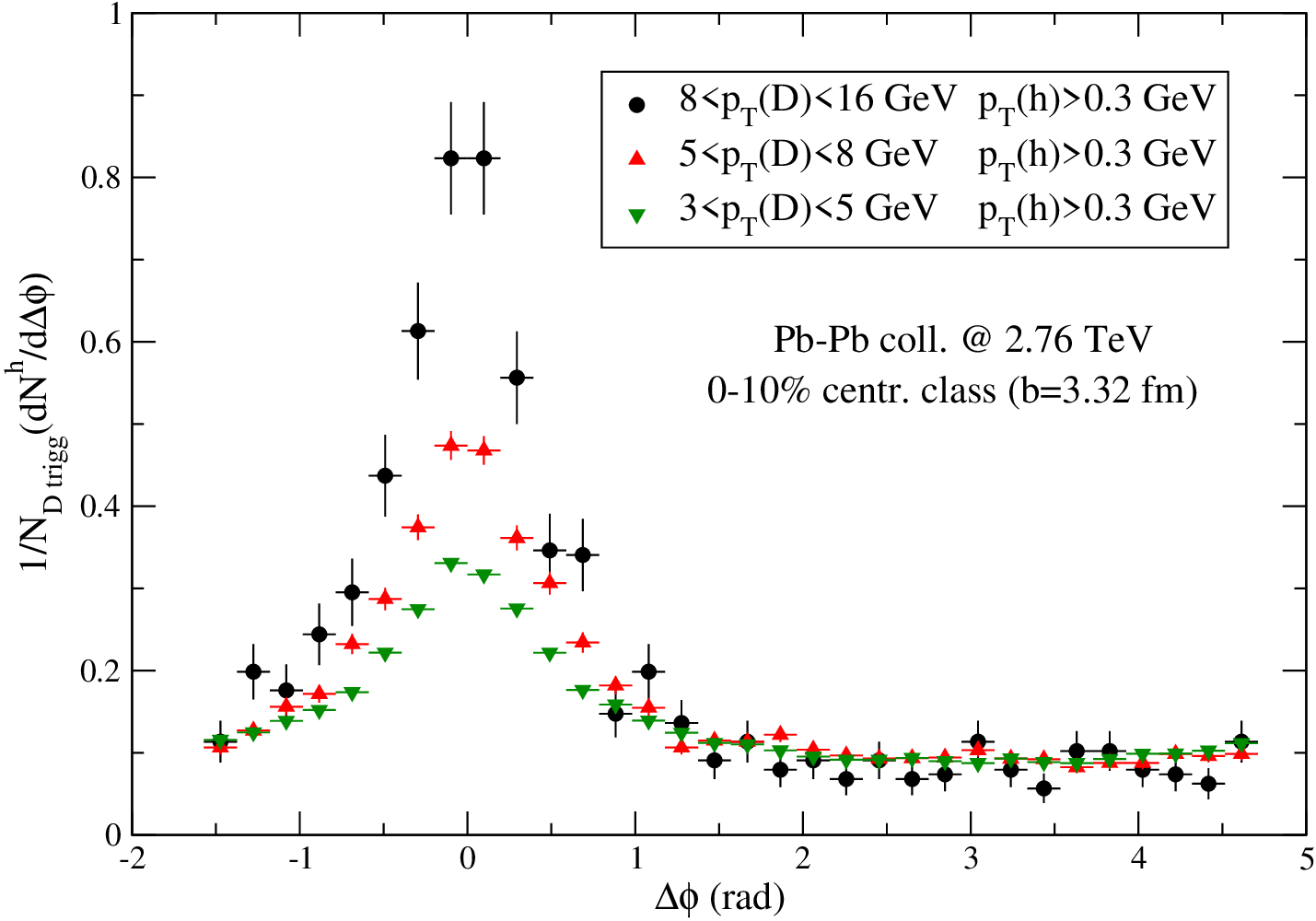}
\includegraphics[clip,width=0.48\textwidth]{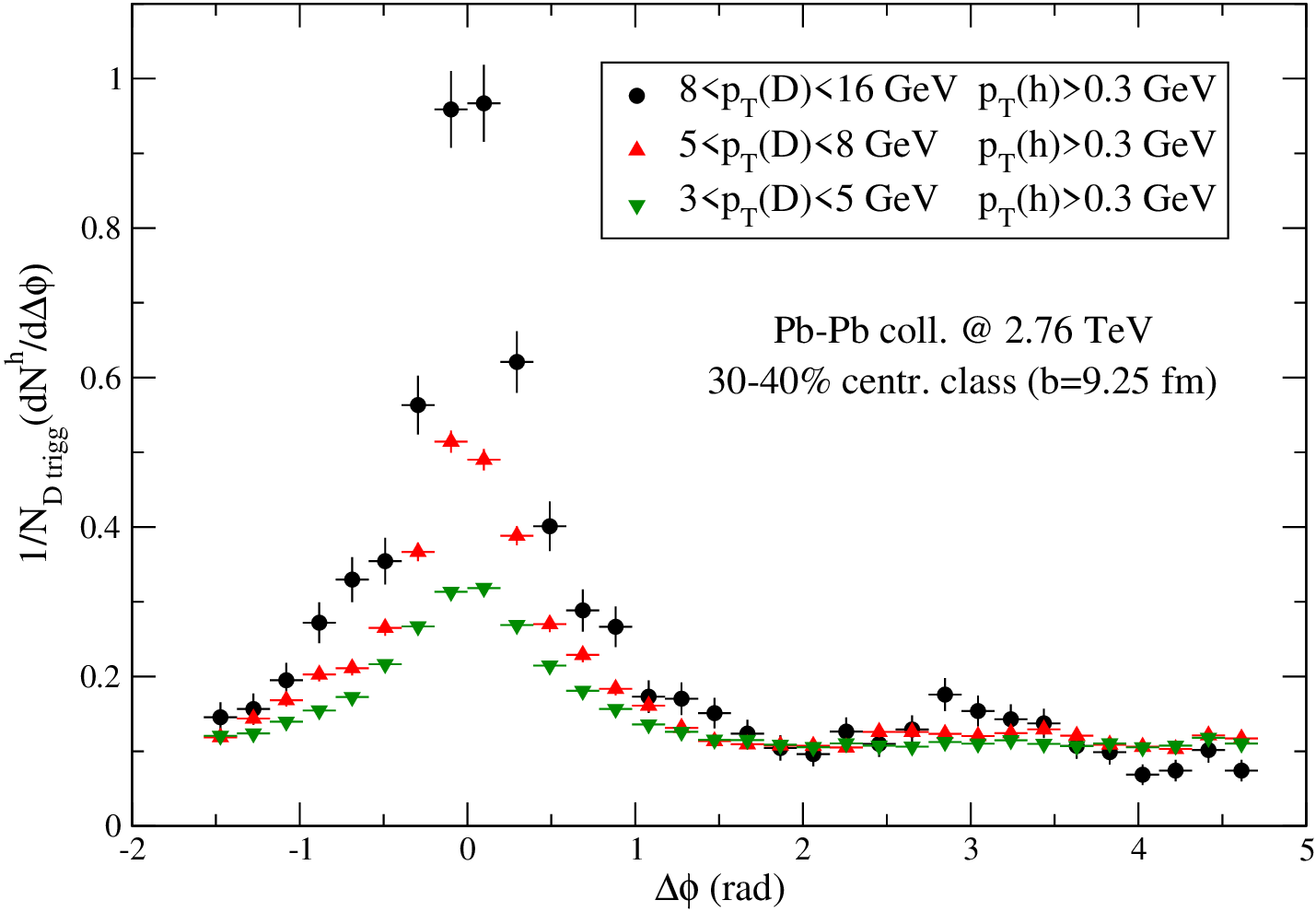}
\caption{$D\!-\!h$ correlations in Pb-Pb collisions at $\sqrt{s_{\rm NN}}\!=\!2.76$ TeV for various $p_T$ cuts on the trigger particle and different centrality classes: $0\!-\!10\%$ (left panel) and $30\!-\!40\%$ (right panel).}\label{Dhcorr}
\end{center}
\end{figure}
\begin{figure}[!h]
\begin{center}
\includegraphics[clip,width=0.48\textwidth]{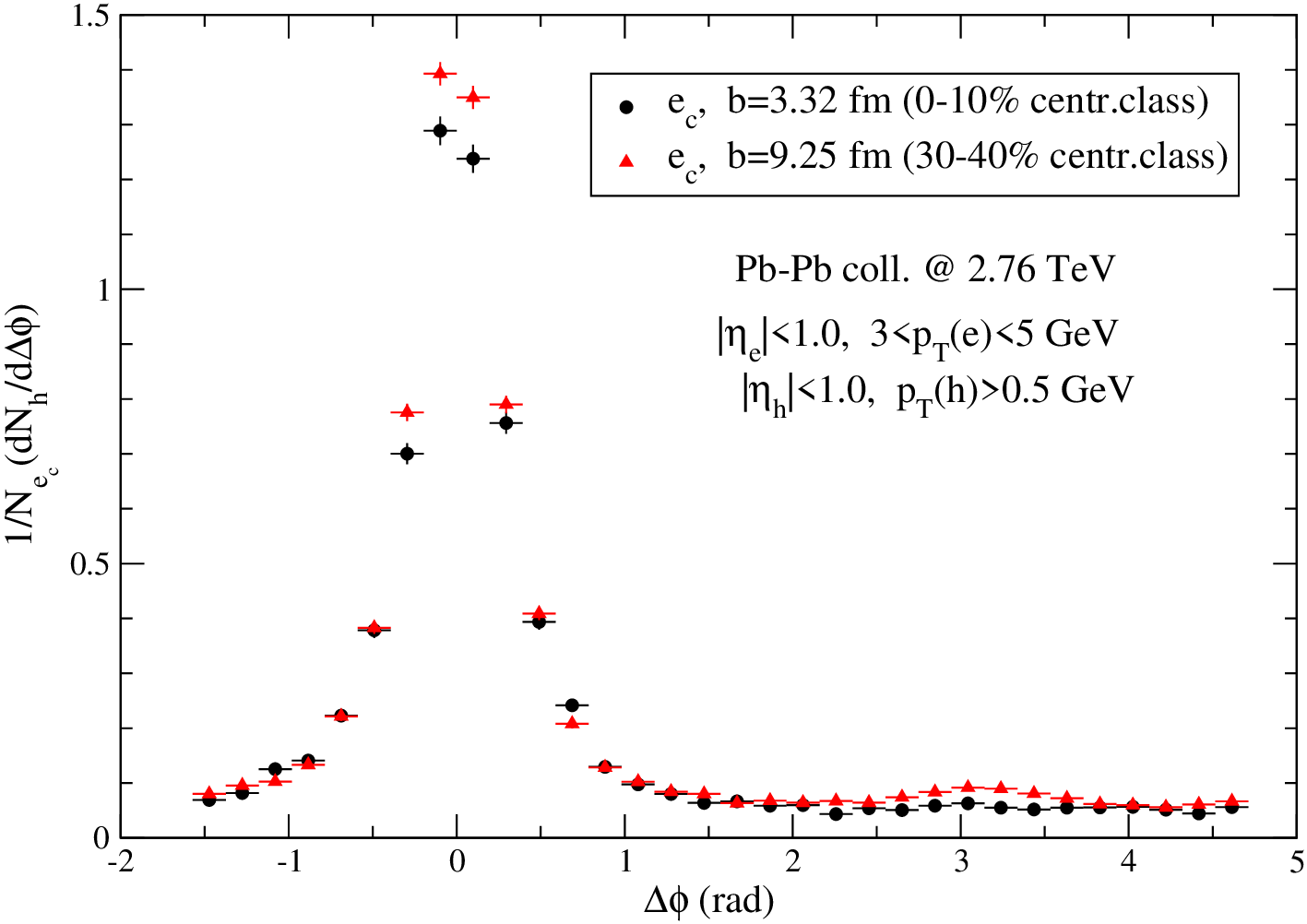}
\includegraphics[clip,width=0.48\textwidth]{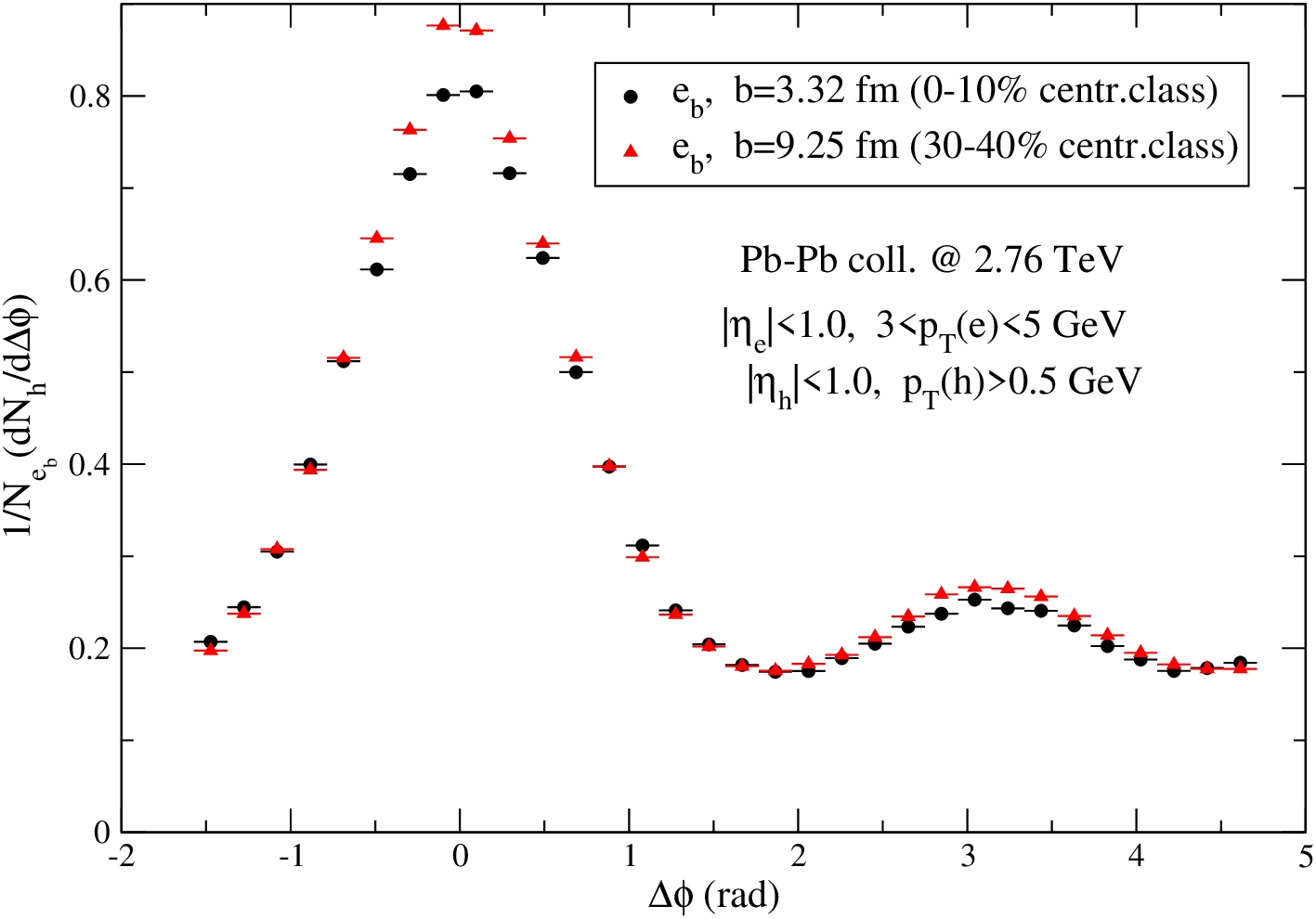}
\caption{$e\!-\!h$ correlations in Pb-Pb collisions at $\sqrt{s_{\rm NN}}\!=\!2.76$ TeV for various centrality classes: $0\!-\!10\%$ (black circles) and $30\!-\!40\%$ (red triangles). The separate charm (left panel) and beauty (right panel) results are shown.}\label{ehcorr} 
\end{center}
\end{figure}
Finally we decide to address, although at the limit of the present experimental capabilities,
more differential observables like angular correlations between heavy flavour particles (or their decay products) and the charged hadrons produced in the same collision. In Figs.~\ref{Dhcorr} and \ref{ehcorr} we display then our results for $D\!-\!h$ and $e\!-\!h$ correlations. All figures are obtained with weak-coupling HTL transport coefficients in the QGP phase. In general one observes a strong suppression of the away-side peak around $\Delta\phi\!=\!\pi$. Depending on the cuts imposed on the trigger particles (either $D$-mesons or heavy-flavour decay electrons) and on the associated hadrons this can be mostly due either to the energy loss (moving particles below the $p_T$-cut) or to the angular decorrelation (moving particles away from $\Delta\phi\!=\!\pi$) of the parent heavy quark. Concerning the $e\!-\!h$ correlations we have plotted separately the charm and beauty contributions. While in the case of electrons from charm decays we have always found an almost complete suppression of the away-side peak independently on the centrality of the collision and on the kinematic cuts, beauty decay electrons turn out to be less decorrelated, allowing one in principle to extract more information on the heavy-quark interaction with the medium.

In summary, the new in-medium fragmentation model for heavy quark hadronization we developed, once interfaced to our POWLANG transport setup, allows us to get results for the $D$ meson $R_{AA}$ and $v_2$ in better agreement with the experimental data. We have also tried to extend our predictions to more differential observables, like heavy flavour angular correlations: this has to be considered just an exploratory study, leaving the issue of the information one can extract from such a kind of measurements for future work.

\end{document}